\documentclass[aps,prl,twocolumn,superscriptaddress,showpacs,floatfix,nobibnotes]{revtex4}
\usepackage{subfigure}
\usepackage{epsfig}
\usepackage{epstopdf}
\usepackage{color}
\usepackage{graphicx}
\usepackage{longtable}
\usepackage{CJK}
\usepackage{amsmath,bbold}
\usepackage{float}
\newcommand{\blue}[1]{{ \color{blue} #1 }}
\ifx\pdfoutput\undefined
\usepackage[dvips]{graphicx}
\usepackage[dvipdfm,
  pdfstartview=FitH,
     CJKbookmarks=true,
          bookmarksnumbered=true,
          bookmarksopen=true,
         colorlinks=true,
          pdfborder=002,
          citecolor=blue%
           ]{hyperref}
\AtBeginDvi{} 
\else
\usepackage[
          pdfstartview=FitH,
          CJKbookmarks=true,
          bookmarksnumbered=true,
           bookmarksopen=ture,
           colorlinks=true,
           citecolor=blue
           ]{hyperref}
\newcommand{\red}[1]{{ \color{red} #1 }}

\usepackage{graphicx}
\usepackage{longtable}

\begin{document}

\title{The Heine-Stieltjes correspondence and a new angular \\
momentum projection for many-particle systems}

\author{Feng Pan}
\affiliation{Department of Physics, Liaoning Normal University,
Dalian 116029, China}\affiliation{Department of Physics and
Astronomy, Louisiana State University, Baton Rouge, LA 70803-4001,
USA}\affiliation{School of Mathematics and Physics, The University of Queensland,
Brisbane, Qld 4072, Australia}

\author{Bo Li}
\affiliation{Department of Physics, Liaoning Normal University,
Dalian 116029, China}

\author{Yao-Zhong Zhang}
\affiliation{School of Mathematics and Physics, The University of Queensland,
Brisbane, Qld 4072, Australia}

\author{Jerry P. Draayer}
\affiliation{Department of Physics and Astronomy, Louisiana State
University, Baton Rouge, LA 70803-4001, USA}
\date{\today}

\begin{abstract}
A new angular momentum projection for systems of particles with arbitrary spins
is formulated based on the Heine-Stieltjes correspondence,
which can be regarded as the solutions of the mean-field plus
{pairing} model in the strong pairing interaction $G\rightarrow\infty$ limit.
Properties of the Stieltjes zeros of the extended Heine-Stieltjes
polynomials, of which the roots determine the projected states,
and the related Van Vleck zeros are discussed.
The electrostatic interpretation of these zeros {is} presented.
As examples, applications to $n$ nonidentical particles of spin-$1/2$
and to identical bosons or fermions are made to elucidate the procedure and
properties of the Stieltjes zeros and the related Van Vleck zeros.
It is shown that the new angular momentum projection for $n$ identical bosons
or fermions can be simplified with the branching multiplicity formula
of $U(N)\downarrow O(3)$ and the special choices of the parameters used
in the projection. Especially, it is shown that the solutions for identical
bosons can always be expressed in terms of zeros of Jacobi polynomials.
However, unlike non-identical particle systems,
the $n$-coupled states of identical particles are non-orthogonal
with respect to the multiplicity label after the projection.
\end{abstract}


\maketitle\section{I. Introduction}

The angular momentum projection or construction of many-body wavefunctions with
definite total angular momentum from a set of single-particle product states
{has practical value in} quantum many-body physics \cite{low,ring,bie}.
For a few particle systems, the Clebsch-Gordan (CG) coefficients,
3j symbols or Wigner coefficients, can be used straightforwardly
for {this} purpose. However, with increasing {particle numbers,}
the CG couplings become tedious and cumbersome {because with increasing
particle numbers, the number of intermediate angular momentum quantum numbers
the are required to label different states with the same total angular momentum grows
combinatorially.}
In practical applications, the projection technique of L\"{o}wdin has been
one of the most popular \cite{low}. This method uses the angular momentum projection
operator to project a set of single-particle product states into states with
definite total angular momentum, {which requires solution of  the eigenvalue problem}
of the projection operator matrix constructed from the relevant
single-particle product states.
In \cite{bie}, Biedenharn and Louck proposed the Wigner operator
method that combines CG couplings with results from the theory of the symmetric groups.
However, their method can only be worked out explicitly
for $n$ nonidentical particles of spin-$1/2$.
In the case of the nuclear shell model, other
procedures are used to construct states with
definite total angular momentum quantum number $J$. One, called
the $M$ scheme, starts with single-particle product states with good
total angular momentum projection and utilizes a simple subtraction
procedure to extract states with good total angular momentum~\cite{ms},
and another uses direct angular momentum couplings and is usually
referred to as the $J$-coupled scheme for identical particles or the
$JT$-coupled scheme when applied to a proton-neutron system~\cite{js}.
Alternatively, the projection operator constructed in terms
of an integration of {the} product of the rotational group element
and its matrix element (Wigner's D-function) of a given angular momentum
over the Euler angles {can also be used~\cite{ring}, as} or example, in
construction of Elliott basis~\cite{elliott} of $SU(3)\supset SO(3)$
and in the projected shell model calculations~\cite{hara}.
These methods can {all be relatively easily} implemented in computer codes
{designed for their respective} purpose.
Their drawbacks lie in the fact that much CPU time is needed
when the dimension of the subspace spanned by the relevant single-particle product states
{is really large,} especially when the projection operator {is} constructed in terms
of an integration of the  product of the rotational group element
and its matrix element of a given angular momentum
over the Euler angles is used because
the Wheeler-Hill integral involved is difficult to be treated accurately in the code.

Recently, it has been shown that the angular momentum projection
may be realized by solving a set of Bethe ansatz equations~\cite{pan1,Guan}.
The purpose of this work is to show that the Bethe ansatz equations
can be solved relatively easily from zeros of the associated extended Heine-Stieltjes polynomials
from the Heine-Stieltjes correspondence~\cite{pan2,Guan,lerma,bae,links1,zhang}. In Sec. II, we will revisit the Bethe ansatz
method for the angular momentum projection. The Heine-Stieltjes correspondence related to the problem,
together with properties of the Heine-Stieltjes polynomials and their electrostatic interpretation, will be studied
in Sec. III. As an examples, the application to $n$ nonidentical particles with spin-$1/2$ will be
shown in Sec. IV, which is also related to the eigenvalue problem of the pure {pairing} interactions
among valence nucleon pairs over a set of deformed Nilsson orbits, while applications
to systems of identical bosons and fermions will be discussed in Sec. V.
A brief summary will be given in Sec. VI.

\section{II. The Bethe ansatz method for angular momentum projection}

Let $\{J_{\mu}^{\alpha};~\mu = +,~-,~0\}$, where  $\alpha=1,2,\cdots,n$,
be generators of the $\alpha$-th copy of the $SU(2)$ algebra,
which satisfy the commutation relations:
\begin{eqnarray}\label{1}
&[J_{+}^{\alpha},~J_{-}^{\beta}]=2\delta_{\alpha\beta}J_{0}^{\alpha},\, \nonumber &\\
&[J_{0}^{\alpha},~J_{\pm}^{\beta}]=\pm\delta_{\alpha\beta}J_{\pm}^{\alpha},\,
\end{eqnarray}
and
$\vert j_{\alpha}, m_{\alpha}\rangle$
be the corresponding orthonormal basis vectors with angular momentum quantum number $j_{\alpha}$
and quantum number  $m_{\alpha}$ of its third component.
According to
the Bethe ansatz method, one can write $n$-coupled state
with total angular momentum $J=\sum_{\alpha}j_{\alpha}-k$  and $M=J$ as

\begin{equation}\label{2}
\vert \zeta; J,M=J)=J_{-}(x^{(\zeta)}_{1})J_{-}(x^{(\zeta)}_{2})\cdots J_{-}(x^{(\zeta)}_{k})\vert{\rm h.w.}\rangle,
\end{equation}
where
$\vert{\rm h.w.}\rangle=\prod_{\alpha=1}^{n}\vert j_{\alpha}, m_{\alpha}= j_{\alpha}\rangle$
is the SU(2) highest weight state satisfying $J_{+}^{\alpha}\vert{\rm h.w.}\rangle=0$ for any $\alpha$,

\begin{equation}
J_{-}(x^{(\zeta)}_{i})=\sum_{\alpha=1}^{n}{1\over{x^{(\zeta)}_{i}-\epsilon_{\alpha}}}J_{-}^{\alpha},
\end{equation}
in which the parameters $\{\epsilon_{\alpha}\}$
can be any set of unequal numbers, and $\zeta$ is used to distinguish different $n$-coupled
states with the same angular momentum $J$. Because (\ref{2})
is the highest weight state of the angular momentum $J$, it should satisfy
the condition

\begin{eqnarray}\label{3}
&J_{+}\vert \zeta; J,M=J)=\, \nonumber &\\
&J_{+}J_{-}(x^{(\zeta)}_{1})J_{-}(x^{(\zeta)}_{2})\cdots J_{-}(x^{(\zeta)}_{k})\vert{\rm h.w.}\rangle=0,\,
\end{eqnarray}
where $J_{+}=\sum_{\alpha}J_{+}^{\alpha}$.
By using the commutation relations (\ref{1}), (\ref{3}) requires
that the Bethe ansatz equations (BAEs)

\begin{equation}\label{4}
\sum_{\alpha=1}^{n}{2j_{\alpha}\over{x^{(\zeta)}_{i}-\epsilon_{\alpha}}}
-\sum_{t=1(\neq i)}^{k}{2\over{x^{(\zeta)}_{i}-x^{(\zeta)}_{t}}}=0
\end{equation}
must be satisfied for $i=1,2,\cdots,k$. It is clear that
the multiplicity label $\zeta=1,2,\cdots,d(n,k)$ in (\ref{2}) is taken to be the label of
different solutions $\{x^{(\zeta)}\}$ of Eq. (\ref{4}).
It can be verified \cite{pan1,Guan} that the number of solutions $d(n,k)$ of Eq. (\ref{4}) equals exactly
to the multiplicity in the reduction $j_{1}\otimes j_{2}\otimes\cdots\otimes j_{n}\downarrow J$,
which can be calculated by

\begin{equation}\label{5}
d(n,k)=\eta(n,k)-\sum^{k-1}_{\mu=0}d(n,\mu),
\end{equation}
where

\begin{equation}
\eta(n,k)=\sum^{2j_{1}}_{\mu_{1}=0}\cdots\sum^{2j_{n}}_{\mu_{n}=0}\delta_{q,k},
\end{equation}
in which $q=\sum^{n}_{i=1}\mu_{i}$.
From Eqs. (6) and (7), the multiplicity $d(n,k)$ can be calculated recursively
from $d(n,0)=1$.

Once the solutions $\{x^{(\zeta)}_{1},\cdots,x^{(\zeta)}_{k}\}$ are obtained
from Eq. (\ref{4}), the $n$-coupled state with any $M$ can be expressed
in the standard way as

\begin{equation}
\vert \zeta; J,M)
=\sqrt{(J+M)!\over{(2J)!(J-M)!}}J_{-}^{J-M}\vert \zeta; J,J),
\end{equation}
where $\vert \zeta; J,J)$ is given by Eq. (2).

Because the uncoupled basis vectors $\{\vert j_{\alpha}, m_{\alpha}\rangle\}$
are orthonormal, substituting (3) into (2), one can find
that the unnormalized angular momentum multi-coupling coefficient
is given by

\begin{eqnarray}
&(j_{1},j_{1}-\mu_{1};\cdots;j_{n},j_{n}-\mu_{n}\vert \zeta; J,J)=\, \nonumber &\\
&S^{(k)}(\beta_{1}^{\mu_{1}},\cdots,\beta_{n}^{\mu_{n}})
\prod^{n}_{i=1}\sqrt{(2j_{i})!\mu_{i}!\over{(2j_{i}-\mu_{n})!}},\,
\end{eqnarray}
where the condition $\sum^{n}_{\alpha}\mu_{\alpha}=k$ must be satisfied,
$S^{(k)}(\beta_{1}^{\mu_{1}},\cdots,\beta_{n}^{\mu_{n}})$
is the $k\times n$-variable symmetric function, in which
$\beta_{\alpha}^{\mu_{\alpha}}$ is the shorthand notation
of taken $\mu_{\alpha}$ variables $\{\beta_{i_{1},\alpha},\cdots,\beta_{i_{\mu_{\alpha}},\alpha}\}$
with $i_{1}\neq i_{2}\neq\cdots\neq i_{\mu_{\alpha}}$
from $\{\beta_{1,\alpha},\cdots,\beta_{k,\alpha}\}$, and

\begin{equation}
\beta_{i,\alpha}={1\over{x^{(\zeta)}_{i}-\epsilon_{\alpha}}}.
\end{equation}
When $n=2$ and $k=2$, for example, we have
$S^{(2)}(\beta_{1}^{2})=\beta_{11}\beta_{12}$,
$S^{(2)}(\beta_{1},\beta_{2})=\beta_{11}\beta_{22}
+\beta_{21}\beta_{12}$, and $S^{(2)}(\beta_{2}^{2})=\beta_{12}\beta_{22}$.
The normalized angular momentum multi-coupling coefficient
is

\begin{eqnarray}
&\langle j_{1},j_{1}-\mu_{1},\cdots,j_{n},j_{n}-\mu_{n}\vert \zeta; J,J\rangle=\, \nonumber &\\
&(j_{1},j_{1}-\mu_{1},\cdots,j_{n},j_{n}-\mu_{n}\vert \zeta; J,J)/{\cal N},\,
\end{eqnarray}
where
\begin{equation}
{\cal N}=({\sum_{\mu_{1}...\mu_{n}}}(j_{1},j_{1}-\mu_{1},\cdots,j_{n},j_{n}-\mu_{n}\vert \zeta; J,J)^2
)^{{1\over{2}}},
\end{equation}
in which the summation should be restricted by $\sum^{n}_{\alpha}\mu_{\alpha}=k$.

In comparison to the traditional projection methods~\cite{low,ring,js,ms},
the Bethe ansatz method for angular momentum projection is more efficient,
which only needs to solve $k$-coupled algebraic BAEs.
However, one must solve $d(n,k)$-dimensional matrix eigenvalue problem
in the traditional projection methods. The dimension $d(n,k)$ increases
with the increasing of $n$ and $k$ in a non-polynomial way as shown
in (\ref{5}). Therefore, the Bethe ansatz method
for angular momentum projection is advantageous
if there is a simple way to solve the BAEs (\ref{4}).

\section{III. The Heine-Stieltjes correspondence}

It has been shown that the BAEs similar to
those shown in (\ref{4}) may be solved
from zeros of the corresponding extended
Heine-Stieltjes polynomials~\cite{pan2,Guan,lerma,bae,links1,zhang}.
Through the Heine-Stieltjes correspondence~\cite{pan2,Guan},
for the BAEs (\ref{4}), one may consider the following
second-order Fuchsian equation:
\begin{equation}\label{6}
A_{n}(x)y^{\prime\prime}_{k}(x)+B_{n-1}(x)y^{\prime}_{k}(x)-V_{n-2}(x)y_{k}(x)=0,
  \end{equation}
where
$A_{n}(x)=\prod_{\alpha=1}^{n}(x-\epsilon_{\alpha})$
is a polynomial of degree $n$,
the polynomial  $B_{n-1}(x)$ is given as
\begin{equation}
B_{n-1}(x)/A_{n}(x)=-\sum^{n}_{\alpha=1}{2j_{\alpha}\over{x-\epsilon_{\alpha}}}\, ,
\end{equation}
and $V_{n-2}(x)$ is
called Van Vleck polynomial~\cite{Szeg} of degree
$n-2$, which are determined according to Eq. (\ref{6}).
Let $\{x_{i},~i=1,2,\cdots,k\}$ be zeros of the extended
Heine-Stieltjes polynomial $y_{k}(x)$, which are often called
Stieltjes zeros. We may write
$y_{k}(x)=\prod_{i=1}^{k}(x-x_{i})$.
At any zero $x_{i}$ of $y_{k}(x)$ , there is the identity

\begin{equation}
{y^{\prime\prime}_{k}(x_{i})\over{y^{\prime}_{k}(x_{i})}}
=\sum^{k}_{t=1 (t\neq i)}{2\over{x_{i}-x_{t}}}.
\end{equation}
It is obvious that, at any zero $x_{i}$ of $y_{k}(x)$ , (\ref{6}) results in
the BAEs (\ref{4}). Generally, we also have

\begin{equation}\label{7}
{y^{\prime\prime}_{k}(x)\over{y_{k}(x)}}=\sum_{1\leq i<t\leq k}{2\over{(x-x_{i})(x-x_{t})}}=\, \nonumber
\end{equation}
\begin{equation}
\sum_{1\leq i\neq t\leq k}{2\over{x-x_{i}}}{1\over{(x_{i}-x_{t})}}, \,
\end{equation}
\begin{equation}\label{8}
{y^{\prime}_{k}(x)\over{y_{k}(x)}}=\sum_{i=1}^{k}{1\over{x-x_{i}}}.\,
\end{equation}
Substituting (\ref{7}) and (\ref{8}) into (\ref{6}), we have

\begin{equation}\label{9}
V^{(\zeta)}_{n-2}(x)=
A_{n}(x)\times \, \nonumber
\end{equation}
\begin{equation}
\sum_{i=1}^{k}{1\over{x-x^{(\zeta)}_{i}}}\left(
\sum_{t\neq i}{2\over{x^{(\zeta)}_{i}-x^{(\zeta)}_{t}}}-
\sum_{\alpha}{2j_{\alpha}\over{x-\epsilon_{\alpha}}}\right).
\end{equation}
By using the BAEs (\ref{4}), (\ref{9}) becomes

\begin{equation}\label{10}
V^{(\zeta)}_{n-2}(x)=
A_{n}(x)\sum_{\alpha=1}^{n}{1\over{x-\epsilon_{\alpha}}}\left(
\sum_{i=1}^{k}{2j_{\alpha}\over{x^{(\zeta)}_{i}-\epsilon_{\alpha}}}\right).
\end{equation}
Eq. (\ref{10}) shows that zeros $\{\bar{x}^{(\zeta)}_{l};~l=1,2,\cdots,n-2\}$ of the Van Vleck polynomial $V^{(\zeta)}_{n-2}(x)$ related
to the $\zeta$-th extended Heine-Stieltjes polynomial $y^{(\zeta)}_{k}(x)=\prod_{i=1}^{k}(x-x^{(\zeta)}_{i})$
are determined by

\begin{equation}\label{11}
\sum_{\alpha=1}^{n}{1\over{\bar{x}^{(\zeta)}_{l}-\epsilon_{\alpha}}}\left(
\sum_{i=1}^{k}{2j_{\alpha}\over{\epsilon_{\alpha}}-x^{(\zeta)}_{i}}\right)=0.
\end{equation}
$\{\bar{x}^{(\zeta)}_{l};~l=1,2,\cdots,n-2\}$ are called Ven Vleck zeros
related to the $\zeta$-th extended Heine-Stieltjes polynomial $y^{(\zeta)}_{k}(x)$.
Once the Van Vleck zeros are obtained from Eq. (\ref{11}),
$V^{(\zeta)}_{n-2}(x)$ can be expressed explicitly as

\begin{equation}\label{12}
V^{(\zeta)}_{n-2}(x)=c_{n,k}\prod^{n-2}_{l=1}(x-\bar{x}^{(\zeta)}_{l}),
\end{equation}
where $c_{n,k}$ is a constant depending on $n$, $k$, and
the parameters $\epsilon_{\alpha}$ ($\alpha=1,\cdots,n$).

If $\epsilon_{\alpha}$ ($\alpha=1,\cdots,n$) are chosen to be real,
according to the Stieltjes results~\cite{Szeg}, the electrostatic interpretation of the location of zeros
of the extended Heine-Stieltjes polynomial $y_{k}(x)$ may be stated as follows. Put $n$ negative fixed charges
$-j_{\alpha}$  at $\epsilon_{\alpha}$ for $\alpha=1,\cdots,n$ along a real line, respectively,
and allow $k$ positive unit charges to move freely on the two dimensional complex plane.
Therefore, up to a constant, the total energy functional  may
be written as

\begin{equation}
U(x_1, x_2,\cdots, x_k)=\, \nonumber
\end{equation}
\begin{equation}
\sum_{i=1}^{k}\sum_{\alpha}^{n}j_{\alpha}\ln\vert x_{i}-\epsilon_{\alpha}\vert-
\sum_{1\leq i\neq t\leq k}\ln\vert x_{i}-x_{t}\vert.
\end{equation}
The BAEs given in Eq. (\ref{4}) imply that there are
$d(n,k)$ different configurations for the position of the $k$
positive charges  $\{x_{1}^{(\zeta)},\cdots, x_{k}^{(\zeta)}\}$
with $\zeta=1,2,\cdots,d(n,k)$, corresponding
to global minimums of the total energy.

Similarly, let

\begin{equation}\label{13}
\rho^{(\zeta)}_{\alpha}(k)=2j_{\alpha}\sum_{i=1}^{k}{1\over{\epsilon_{\alpha}}-x^{(\zeta)}_{i}},
\end{equation}
which is now called Van Vleck charges related to the zeros of the $\zeta$-th
extended Heine-Stieltjes polynomial $y_{k}^{(\zeta)}(x)$. Put $n$ Van Vleck charges $\rho^{(\zeta)}_{\alpha}(k)$
at positions $\epsilon_{\alpha}$ for $\alpha=1,\cdots,n$ along a real line, respectively,
and allow one unit charge to move freely on the two dimensional complex plane.
Eq. (\ref{11}) provides $n-2$ possible equilibrium positions  $\{\bar{x}^{(\zeta)}_{l};~l=1,2,\cdots,n-2\}$
of the unit moving charge for the electrostatic system.

Let

\begin{equation}
\Lambda(x)={y^{\prime}_{k}(x)\over{y_{k}(x)}}=\sum_{i=1}^{k}{1\over{x-x_{i}}}.\,
\end{equation}
As shown in \cite{far}, $\Lambda(x)$ satisfies the Riccati type equation

\begin{equation}\label{14}
\Lambda^{\prime}(x)+\Lambda^{2}(x)+\sum_{i=1}^{k}\sum_{\alpha}^{n}{2j_{\alpha}\over{(x-x_{i})(\epsilon_{\alpha}-x_{i})}}=0
\end{equation}
in this case. The Van Vleck charges $\rho_{\alpha}$ can be expressed as

\begin{equation}
\rho_{\alpha}=2j_{\alpha}\Lambda(\epsilon_\alpha).
\end{equation}
There are a series of high order differential equations~\cite{far} for $\Lambda(\epsilon_\alpha)$.
For example, the lowest order one is

\begin{equation}\label{15}
(1-2j_{\beta})\Lambda^{\prime}(\epsilon_{\beta})+\Lambda^{2}(\epsilon_{\beta}) \nonumber
\end{equation}
\begin{equation}
+\sum_{\alpha\neq\beta}{2j_{\alpha}}
{\Lambda(\epsilon_{\beta})-\Lambda(\epsilon_{\alpha})\over{\epsilon_{\alpha}-\epsilon_{\beta}}}=0.
\end{equation}

It seems that the solutions of $\Lambda(\epsilon_\alpha)$ of Eq. (\ref{15})
or from a series of high order differential equations
can be used to determine  Van Vleck zeros, and eventually solve
the BAEs (\ref{4}) as shown in ~\cite{far}. It should be noted
that the solutions of $\{\Lambda(\epsilon_\alpha)\}$ from those Riccati
type equations only depend on the parameters $\{\epsilon_{\alpha}\}$
and $\{j_{\alpha}\}$,
but do not explicitly depend on $k$ and $\zeta$. Therefore,
the solutions of $\{\Lambda(\epsilon_\alpha)\}$ from those Riccati
type equations are numerous. One should try to
search for a set of solution $\{\Lambda(\epsilon_\alpha)\}$ corresponding to  specific $k$ and $\zeta$
from solutions with all possible $k$ and $\zeta$ obtained from Eq. (\ref{15}) ,
which explains why the other
expressions of $\{\Lambda(\epsilon_\alpha)\}$ in terms
of symmetric functions of $\{x_{1},\cdots,x_{k}\}$
should also be used to solve the problem~\cite{far}.

In order to avoid the previous mentioned ambiguity, in the following,
we insist on using the method outlined in~\cite{pan2}.
We write
\begin{equation}\label{16}
y_{k}(x)=\sum_{j=0}^{k}a_{j}x^{j},~~V_{n-2}(x)=\sum_{j=0}^{n-2}b_{j}x^{j},
\end{equation}
where $\{a_{j}\}$ and $\{b_{j}\}$ are the expansion coefficients to be determined.
Substitution of  (\ref{16}) into (\ref{6}) yields two matrix equations.
By solving these two matrix equations, we can obtain the solutions of $\{a_{j}\}$ and $\{b_{j}\}$
for given $k$.
Since there {is freedom} to choose the parameters $\{\epsilon_{\alpha};~\alpha=1,\cdots,n\}$,
we find the following parameter settings {to be a simple and convenient choice} due
to the fact that there is an additional reflection symmetry in (\ref{4}):

\begin{equation}
\left\{
\begin{tabular}{c}
$\epsilon_{\alpha}=-(p+1-\alpha)~{\rm for}~\alpha\leq p$, \\
$\epsilon_{\alpha+p}=\alpha~~{\rm for}~\alpha\geq1$,~~~~~~~~~~~\\
\end{tabular}\right.
\end{equation}
when $n=2p$, and

\begin{equation}
\left\{
\begin{tabular}{c}
$\epsilon_{\alpha}=-(p+1-\alpha)~{\rm for}~\alpha\leq p$, \\
$\epsilon_{p+1}=0$,~~~~~~~~~~~~~~~~~~~~~~~~~~\\
$\epsilon_{\alpha+p+1}=\alpha~~{\rm for}~\alpha\geq1$,~~~~~~~~\\
\end{tabular}\right.
\end{equation}
when $n=2p+1$. With such \red{a} choice, in addition to the $S_{k}$ permutation
symmetry among indices $i=1,\cdots,k$ of $\{x_{1},\cdots,x_{k}\}$,
the Stieltjes zeros $\{x_{i}\}$ have the following additional
reflection symmetries: (i) If $\{x_{1},\cdots,x_{k}\}$ is a set of Stieltjes zeros,
$\{-x_{1},\cdots,-x_{k}\}$ is another set.
(ii)When $n$ is even, there are many sets of solutions with $\{x_{1}=-x_{2},x_{3}=-x_{4},
\cdots,x_{k-1}=-x_{k}\}$ when $k$ is even, and $\{x_{1}=-x_{2},x_{3}=-x_{4},
\cdots,x_{k-2}=-x_{k-1},x_{k}=0\}$ when $k$ is odd.
When $n$ is odd, there are many sets of solutions with $\{x_{1}=-x_{2},x_{3}=-x_{4},
\cdots,x_{k-1}=-x_{k}\}$ when $k$ is even. Solutions satisfying
property (ii) are self-reflectional.
Property (i) is strong, namely such {pairs} of solutions always exist, which
is obvious with the substitutions of  $x_{i}$ with $-x_{i}$
in Eq. (\ref{4}) for $i=1,\cdots,k$. However, property (ii)
only applies to a subset of solutions, namely there are
other sets of solutions which may not follow property (ii).
One can verify that {the} substitution of $\{x_{1}=-x_{2},x_{3}=-x_{4},
\cdots,x_{k-1}=-x_{k}\}$ for $k$ even or $\{x_{1}=-x_{2},x_{3}=-x_{4},
\cdots,x_{k-2}=-x_{k-1},x_{k}=0\}$ for $k$ odd into Eq. ({\ref{4})
indeed yields $k$ consistent equations when $n$ is even, which implies
that  $\{x_{1}=-x_{2},x_{3}=-x_{4},
\cdots,x_{k-1}=-x_{k}\}$ for $k$ even or $\{x_{1}=-x_{2},x_{3}=-x_{4},
\cdots,x_{k-2}=-x_{k-1},x_{k}=0\}$ for $k$ odd is possible solutions
when $n$ is even. For odd $n$ cases, self-reflectional solutions
only exist when $k$ is even.
Because the parameters chosen satisfy the
interlacing condition $\epsilon_{1}<\cdots<\epsilon_{n}$,
zeros of $y_{k}(x)$ may be arranged to satisfy
the interlacing condition,
${\bf Re}(x_{1})\leq{\bf Re}(x_{2})\leq\cdots\leq{\bf Re}(x_{k})$, where
${\bf Re}(x_{i})$ lies in one of the $n-1$ intervals
$(\epsilon_{1},\epsilon_{2})$,
$\cdots$, $(\epsilon_{n-1},\epsilon_{n})$, in which
the equality is only possible when the adjacent zeros
are complex conjugate with each other. When two
zeros are conjugate with each other with $x_{i}=x_{i+1}^{*}$,
it is obvious that ${\bf Re}(x_{i})$ and ${\bf Re}(x_{i+1})$
are in the same interval  $(\epsilon_{\alpha},\epsilon_{\alpha+1})$.
The number of different such allowed configurations gives the possible solutions
of $y_{k}(x)$ and the corresponding $V_{n-2}(x)$.
Therefore, these properties are much helpful to
simplify the problem and in search for solutions of (\ref{4}).

\section{IV. Application to systems with nonidentical spin-${\bf 1/2}$ particles}

Generally, the Bethe ansatz method for angular momentum projection
with the Heine-Stieltjes correspondence shown in previous sections
can be applied to construct state with definite angular momentum
from a set of uncoupled single-particle states of both nonidentical-
and identical-particle systems. Because identical-particle systems
have additional permutation symmetries, namely symmetric
among identical bosons or antisymmetric among identical
fermions with respect to the single-particle coordinate permutations,
the procedure outlined in previous sections can be simplified.
Such simplifications and applications will be shown in the next section.
In this section, we only focus on a nonidentical-particle case, in which
we strictly follow the method described previously because
no further simplification can be made for nonidentical-particle systems.

As the simplest but nontrivial example, we consider $n$ nonidentical particles of spin-$1/2$,
which was previously studied by Louck and Biedenharn using the pattern calculus
with the Yamanocchi symbol of an irrep of $S_{n}$ as the upper pattern used to label
the outer multiplicity of $SU(2)\times\cdots\times SU(2)\downarrow SU(2)$,
and the $SU(2)$ basis of the same irrep as the lower pattern~\cite{bie,lb}.
This is the only case that can be solved analytically by using the Wigner operator
method. However, as shown in
\cite{bie}, the construction of coupled state with definite angular momentum
for nonidentical particles of arbitrary spin can never be expressed analytically
by using the pattern calculus, principally because of unsolved problems relating
to the upper patterns.  Specifically, by using the pattern calculus,
the $n$-coupled state with total angular momentum $J$ of spin-$1/2$ system
may be written as~\cite{bie,lb}

$$\vert (i_1\cdots i_{n});JM\rangle=$$$$
\sum_{k_1\cdots k_n}
\left\langle
\begin{array}{l}
2J~~0\\
J+M\\
\end{array}\right\vert
\left\langle
\begin{array}{l}
~~i_1\\
1~~0\\
~~k_n\\
\end{array}\right\rangle\cdots
\left\langle
\begin{array}{l}
~~i_n\\
1~~0\\
~~k_1\\
\end{array}\right\rangle
\left\vert
\begin{array}{l}
0~~0\\
~0\\
\end{array}\right\rangle\times$$
\begin{equation}
\prod_{i=1}^{n}\vert
{1\over{2}},k_i-{1\over{2}}
\rangle,
\end{equation}\label{17}
where $(i_{1}\cdots i_{n})$ with  $i_{s}=0$ or $1$ for $s=1,\cdots,n$,
is used as the multiplicity label, the sum should be restricted with $k_{i}=0$ or $1$ for
$i=1,\cdots,n$, and the expansion coefficient

\begin{equation}\label{18}
\left\langle
\begin{array}{l}
2J~~0\\
J+M\\
\end{array}\right\vert
\left\langle
\begin{array}{l}
~~i_1\\
1~~0\\
~~k_n\\
\end{array}\right\rangle\cdots
\left\langle
\begin{array}{l}
~~i_n\\
1~~0\\
~~k_1\\
\end{array}\right\rangle
\left\vert
\begin{array}{l}
0~~0\\
~0\\
\end{array}\right\rangle\end{equation}
should be calculated consecutively with

$$\left\langle
\begin{array}{l}
~~1\\
1~~0\\
~~1\\
\end{array}\right\rangle
\left\vert
\begin{array}{l}
2j~~0\\
~j+m\\
\end{array}\right\rangle
=\left(
{j+m+1\over{2j+1}}\right)^{1/2}
\left\vert
\begin{array}{l}
2j+1~~0\\
~j+m+1\\
\end{array}\right\rangle,$$

$$\left\langle
\begin{array}{l}
~~1\\
1~~0\\
~~0\\
\end{array}\right\rangle
\left\vert
\begin{array}{l}
2j~~0\\
~j+m\\
\end{array}\right\rangle
=\left(
{j-m+1\over{2j+1}}\right)^{1/2}
\left\vert
\begin{array}{l}
2j+1~~0\\
~j+m\\
\end{array}\right\rangle,$$

$$\left\langle
\begin{array}{l}
~~0\\
1~~0\\
~~1\\
\end{array}\right\rangle
\left\vert
\begin{array}{l}
2j~~0\\
~j+m\\
\end{array}\right\rangle
=-\left(
{j-m\over{2j+1}}\right)^{1/2}
\left\vert
\begin{array}{l}
2j-1~~0\\
~j+m\\
\end{array}\right\rangle,$$

\begin{equation}\label{19}\left\langle
\begin{array}{l}
~~0\\
1~~0\\
~~0\\
\end{array}\right\rangle
\left\vert
\begin{array}{l}
2j~~0\\
~j+m\\
\end{array}\right\rangle
=\left(
{j+m\over{2j+1}}\right)^{1/2}
\left\vert
\begin{array}{l}
2j-1~~0\\
~j+m-1\\
\end{array}\right\rangle.
\end{equation}

Though the expression of the expansion coefficients shown by
(\ref{18}) is analytic, the evaluation of (\ref{18})
according to the rules shown in (\ref{19}) is still
cumbersome, especially the coefficients for many
permissible upper patterns $(i_1\cdots i_n)$
that may lie in the null space
are zero, which, however, can not be ruled out
beforehand. This is the main drawback in using
the upper pattern to resolve the outer multiplicity
problem of unitary groups~\cite{pandra}.
In contrast, roots of the BAEs (\ref{4})
provide with all possible coupled states
with the same angular momentum $J$ as shown by (\ref{2})
and ({3}), which are mutually orthogonal
with respect to the multiplicity label $\zeta$.
Solutions of (\ref{4}) can be obtained from Eq. (\ref{6})
with the explicit expressions shown in (28).
Moreover, the new angular momentum projection method
outlined in Sec. II is not restricted to systems consisting
of particles with the same spin, but can be applied to
systems consisting of particles with arbitrary spins.

The above example is closely related to the construction
of eigenstates of the pure pairing Hamiltonian in the deformed
Nilsson basis with

\begin{equation}\label{NP}
\hat{H}_{\rm S}=-GS^{+}S^{-},
\end{equation}
where $S^{+}=\sum_{\mu}S^{+}_{\mu}=\sum_{\mu}a^{\dagger}_{\mu\uparrow}a^{\dagger}_{\mu\downarrow}$
and $S^{-}=\left(S^{+}\right)^{\dagger}$, in which
$S^{+}_{\mu}=a^{\dagger}_{\mu\uparrow}a^{\dagger}_{\mu\downarrow}$
($S^{-}_{\mu}=a_{\mu\downarrow}a_{\mu\uparrow}$)
are pair creation (annihilation) operators. The up and down arrows in
these expressions refer to time-reversed states.
For simplicity, we only consider the seniority zero cases.
The eigenstates of (\ref{NP}) can be constructed in the following way~\cite{pan}:
Since each Nilsson level can be occupied at most by one pair due to the
Pauli principle, the local states can be regarded as
quasi-spin-$1/2$ states.  $\vert{1\over{2}},{1\over{2}}\rangle$
stands for one pair state, while $\vert{1\over{2}},-{1\over{2}}\rangle$
stands for no pair state. Then, similar to (\ref{2}),
any allowed total quasi-spin $S$ and $M_{S}=S$ state of $p$ pairs over $n$ Nilsson levels
can be written as

\begin{equation}\label{qs}
\vert\zeta; S, M_{S}=S\rangle={\cal N} S^{-}(x^{(\zeta)}_{1})\cdots S^{-}(x^{(\zeta)}_{t})\vert{\rm h. w.}\rangle
\end{equation}
with $p=n-t$ pairs, where ${\cal N}$ is the normalization constant defined by (12), $S=n/2-t$,
$\vert{\rm h. w.}\rangle\equiv\vert {1\over{2}},{1\over{2}};\cdots;{1\over{2}},{1\over{2}}\rangle$
is the product of $n$ copies of local state with highest weight of quasi-spin-$1/2$,
and

\begin{equation}
S^{-}(x^{(\zeta)}_{i})=\sum_{\mu=1}^{n}{1\over{x^{(\zeta)}_{i}-\epsilon_{\mu}}}S^{-}_{\mu},
\end{equation}
in which the parameters $\{\epsilon_{\mu}\}$
can be any set of unequal numbers, and $\zeta$ is used to distinguish different $n$-coupled
states with the same quasi-spin $S$. The variables $\{x^{(\zeta)}_{1},\cdots, x^{(\zeta)}_{t}\}$
should satisfy

\begin{equation}
\sum_{\mu=1}^{n}{1\over{x^{(\zeta)}_{i}-\epsilon_{\mu}}}
-\sum_{l=1(\neq i)}^{t}{2\over{x^{(\zeta)}_{i}-x^{(\zeta)}_{l}}}=0
\end{equation}
for $i=1,2,\cdots,t$. It is clear that
the multiplicity label $\zeta=1,2,\cdots,d(n,t)$ in (37) is taken to be the label of
different solutions $\{x^{(\zeta)}\}$ of Eq. (37).
It can be verified that the number of solutions $d(n,t)$ of Eq. (37)
equals exactly to the multiplicity in the reduction $j_{1}\otimes j_{2}\otimes\cdots\otimes j_{n}\downarrow J$
with $j_{l}={1\over{2}}$ for $1\leq l\leq n$,
which can be calculated from Eq. (6) with

\begin{equation}
\eta(n,t)=\sum^{1}_{\mu_{1}=0}\cdots\sum^{1}_{\mu_{n}=0}\delta_{q,t}
\end{equation}
for this case, in which $q=\sum^{n}_{i=1}\mu_{i}$.
From Eqs. (6) and (38), the multiplicity $d(n,t)$ can be calculated recursively
with $d(n,0)=1$, which indicates that there are $d(n,t)$
different states with the same quasi-spin $S=n/2-t$.
For this case, there is a closed form of $d(n,t)$ with

\begin{equation}
d(n,t)={(1 + n - 2 t) n!\over{(1 + n - t) (n - t)! t!}},
\end{equation}
which equals exactly to the dimension of the irrep $[n-t,t]$ of the
permutation group $S_{n}$~\cite{bie,pan} and is consistent with
the result obtained from Eqs. (6) and (38).
In this case,  $c_{n,k}$ in the Van Vleck polynomials (21)
can be obtained in solving the corresponding Fuchsian equation (13) with

\begin{equation}
c_{n,k}=-(n-k+1)k.
\end{equation}

Finally, the state with quasi-spin $S$ and any $M_{S}$ can be expressed as

\begin{equation}
\vert \zeta; S,M_{S}\rangle
=\sqrt{(S+M_{S})!\over{(2S)!(S-M_{S})!}}\left(S^{-}\right)^{S-M_{S}}\vert \zeta; S,S\rangle
\end{equation}
for the system with $p=n/2+M_{S}$ pairs.

\begin{table}[htb]
\caption{The multiplicity $d(8,t)$ for  $S=4-t$ and
$d(7,t)$ for $S={7\over{2}}-t$.}
\begin{tabular}{cccccc}\hline\hline
$t$~~&$S=4-t$~~ & $d(8,t)$~~&\vline&$S={7\over{2}}-t$~~&$d(7,t)$\\
\hline
$0$~~&$4$~~&$1$~~&\vline&${7\over{2}}$~~&$1$\\
$1$~~&$3$~~&$7$~~&\vline&${5\over{2}}$~~&$6$\\
$2$~~&$2$~~&$20$~~&\vline&${3\over{2}}$~~&$14$\\
$3$~~&$1$~~&$28$~~&\vline&${1\over{2}}$~~&$14$\\
$4$~~&$0$~~&$14$~~&\vline\\
\hline\hline
\end{tabular}
\end{table}

In order to demonstrate the method and properties
of the zeros outlined in previous sections,
in the following, we display results of the method for
relatively simple cases with $n=8$, $t=1,~4$ and $n=7$, $t=2,~3$ as examples of even and odd $n$ case,
respectively.
The multiplicities $d(8,t)$ with $0\leq t\leq 4$  and
$d(7,t)$ with $0\leq t\leq 3$
are listed in Table I.
With parameters $\{\epsilon_{\alpha}\}$
chosen according to (29) and (30), we find there are exactly $d(n,t)$ different
solutions for given $n$ and $t$ as shown in Table II-V.
For any case, it can be verified that any zero $x^{(\zeta)}_{i}$
of $y^{(\zeta)}_{t}(x)$ indeed lies in one of the $n-1$ intervals
$(\epsilon_{1},\epsilon_{2})$, $\cdots$, $(\epsilon_{n-1},\epsilon_{n})$.
It is obvious that  $y_{1}^{(1)}(x)$ in Table II,
$y^{(1)}_{4}(x),~\cdots,y^{(6)}_{4}(x)$ in Table III,
and  $y^{(1)}_{2}(x),y^{(2)}_{2}(x)$ in Table IV
are self-reflectional. While the solutions in most cases
satisfy the reflection symmetry property (i).
For example, $y^{(7)}_{4}(x)=y^{(8)}_{4}(-x)$, $y^{(9)}_{4}(x)=y^{(10)}_{4}(-x)$,
$y^{(11)}_{4}(x)=y^{(12)}_{4}(-x)$, $y^{(13)}_{4}(x)=y^{(14)}_{4}(-x)$
when $n=8$ and $t=4$. The Van Vleck polynomial
satisfies the same reflection property as that of the corresponding
extended Heine-Stieltjes polynomial. In addition, one can verify that
the Van Vleck zeros of $V^{(\zeta)}_{n-2}(x)$ indeed satisfy
Eq. (20).
With Stieltjes zeros  $\{x_{i}\}$
of $y^{(\zeta)}_{t}(x)$ obtained from Table II-V,
one can verify that the eigenstates (41) are mutually orthogonal with respect to the
multiplicity label $\zeta$:

\begin{equation}
\langle \zeta; S,M_{S}\vert \zeta^{\prime}; S^{\prime},M^{\prime}_{S}\rangle
=\delta_{\zeta,\zeta^{\prime}}\delta_{S,S^{\prime}}\delta_{M_{S},M^{\prime}_{S}}.
\end{equation}

\begin{table*}[htb]
\caption{The extended Heine-Stieltjes Polynomials $y^{(\zeta)}_{1}(x)$ for constructing $S=3$ states with $n=8$ and $t=1$
according to (35) and the corresponding Van Vleck Polynomials $V^{(\zeta)}_{6}(x)$.}
\begin{tabular}{ccc}
\hline\hline
&{The extended  Heine-Stieltjes Polynomials $y^{(\zeta)}_{1}(x)$} &{The Van Vleck Polynomials $V^{(\zeta)}_{6}(x)$} \\
\hline
\hline\\
$\zeta=1$&$x$&$-8(x-3.679)(x-2.59)(x-1.502)(x+1.502)(x+2.59)(x+3.679)$\\\\
$\zeta=2$&$x+3.679$ &$-8 (x-3.679)(x-1.502)(x+1.502)(x+2.59)(x-2.59)x$\\\\
$\zeta=3$&$x-3.679$&$-8(x+3.679)(x+1.502)(x-1.502)(x-2.59)(x+2.59)x$\\\\
$\zeta=4$&$x-2.59$&$-8(x-3.679)(x+3.679)(x-1.502)(x+1.502)(x+2.59)x$\\\\
$\zeta=5$&$x+2.59$&$-8(x+3.679)(x-3.679)(x+1.502)(x-1.502)(x-2.59)x$\\\\
$\zeta=6$&$x+1.502$&$-8(x-3.679)(x-2.59)(x-1.502)(x+2.59)(x+3.679)x$\\\\
$\zeta=7$&$x-1.502$&$-8(x+3.679)(x+2.59)(x+1.502)(x-2.59)(x-3.679)x$\\
\hline\hline
\end{tabular}
\end{table*}

Once the eigenstates (41) of (34) are obtained,
the results can be used for constructing eigenstates
and calculating eigenvalues
of any mean-field plus pairing model by using
the progressive diagonalization scheme as shown
\cite{pan}. Furthermore, Eqs. (5) and (37) can be regarded
as the same BAEs~\cite{rg} in determining solutions of
the mean-field plus the paring model in the
strong pairing interaction $G\rightarrow\infty$ limit
by replacing the parameters $\{\epsilon_{\alpha}\}$
with $\{2\varepsilon_{\alpha}\}$, where $\{\varepsilon_{\alpha}\}$
are single-particle energies in the corresponding orbits of the
mean-field~\cite{Guan}.

\begin{table*}[htb]
\caption{The extended  Heine-Stieltjes Polynomials $y^{(\zeta)}_{4}(x)$ for constructing $S=0$ states with $n=8$ and $t=4$
according to (35) and the corresponding Van Vleck Polynomials $V^{(\zeta)}_{6}(x)$.}
\begin{tabular}{ccc}
\hline\hline
&{The extended  Heine-Stieltjes Polynomials $y^{(\zeta)}_{4}(x)$} &{The Van Vleck Polynomials $V^{(\zeta)}_{6}(x)$} \\
\hline \hline
$\zeta=1$&$(x^2+0.379415)(x^2+10.53874)$&$-20 (x^2-13.0977) (x^2-6.2593)(x^2-1.9184)$\\\\
$\zeta=2$&$(x^2-12.56879) (x^2+0.57829)$ &$-20 (x^2-2.3151)(x^2-4.9901 x+6.4066)(x^2+4.9901 x+6.4066)$\\\\
$\zeta=3$&$(x^2+0.8145)(x^2-5.7144)$&$-20(x^2-12.8761) (x^2 - 2.8274 x+2.1700) (x^2 + 2.8274 x +2.1700)$\\\\
$\zeta=4$&$(x^2-2.23204 )(x^2-12.3149)$&$-20 (x^2+0.5558) (x^2 -
   5.3854 x + 7.4056) (x^2+5.3854 x + 7.4056 )$\\\\
$\zeta=5$&$(x^2+4.82433)(x^2-1.80406)$&$-20 (x^2-13.0201) (x^2-6.1361) (x^2+0.2502)$\\\\
$\zeta=6$&~~$(x^2- 5.07233 x + 6.61367)(x^2+ 5.07233 x + 6.61367)$&$-20 (x^2-12.5294) (x^2-2.2932) (x^2+0.5729)$\\\\
$\zeta=7$&~~$(x^2- 1.2051 x -8.3519) (x^2+ 1.2051 x+0.8179)$&$-20(x^2 + 2.0270x-5.5609)
(x^2 - 4.8900 x +6.1511)\times$\\&&$(x^2 +2.8630 x +2.2150)$\\\\
$\zeta=8$&~~$(x^2+ 1.2051 x-8.3519) ( x^2-1.2051 x+0.8179 )$&
$-20(x^2- 2.0270x-5.5609)(x^2+4.8900 x +6.1511)\times$\\
&&$(x^2- 2.8630x + 2.2150 )$
\\\\
$\zeta=9$&~~$( x^2- 5.01525 x +6.47201) (x^2+ 5.01525 x +5.28038)$&$-20(x^2+ 5.0460 x +5.3283)(x^2 + 0.3380 x + 0.5727)\times$\\
&&$(x^2- 5.3840 x +7.4014)$\\\\
$\zeta=10$&~~$( x^2+ 5.01525 x +6.47201) (x^2- 5.01525 x +5.28038)$&$-20(x^2- 5.0460 x +5.3283)(x^2 - 0.3380 x + 0.5727)\times$\\
&&$(x^2 + 5.3840 x +7.4014)$\\\\
$\zeta=11$&~~$( x^2+ 2.162707 x +3.56055) (x^2- 2.162707 x -5.00227)$&$-20(x^2 + 6.051 x +8.8268)(x^2- 1.3451 x -0.3943)\times$\\
&&$(x^2 - 4.7057 x +5.6852 )$\\\\
$\zeta=12$&~~$(x^2- 2.162707 x + 3.56055)(x^2 + 2.162707 x  -5.00227)$&$-20(x^2 - 6.051 x +8.8268)(x^2+ 1.3451 x -0.3943)\times$\\
&&$(x^2+ 4.7057 x +5.6852 )$\\\\
$\zeta=13$&~~$(x^2+ 1.0751 x +2.9229)(x^2 - 1.0751 x -3.2612)$&$-20 (x^2-3.3149 x-1.0537) (x^2-2.7178 x+2.1494)\times$\\
&&$(x^2+6.0327x+8.7374)$\\\\
$\zeta=14$&~~$(x^2- 1.0751 x +2.9229)(x^2 + 1.0751 x -3.2612)$&$-20 (x^2+3.3149 x-1.0537) (x^2+2.7178 x+2.1494)\times$\\
&&$(x^2-6.0327x+8.7374)$\\
\hline\hline
\end{tabular}
\end{table*}

\begin{table*}[htb]
\caption{The extended  Heine-Stieltjes Polynomials $y^{(\zeta)}_{2}(x)$ for constructing $S=3/2$ states with $n=7$ and $t=2$
according to (35) and the corresponding Van Vleck Polynomials $V^{(\zeta)}_{5}(x)$.}
\begin{tabular}{ccc}
\hline\hline
~&{The extended  Heine-Stieltjes Polynomials $y^{(\zeta)}_{2}(x)$} &{The Van Vleck Polynomials $V^{(\zeta)}_{5}(x)$} \\
\hline
$\zeta=1$&$(x-2.646)(x+2.646)$&$-12(x-1.5275)(x+1.5275)x^3 $\\\\
$\zeta=2$&$(x-1.5275)(x+1.5275)$&$-12(x-2.646)(x+2.646)x^3$\\\\
$\zeta=3$&$x^2- 1.270 x + 0.5564 $&$-12(x-2.63447)(x-1.4683x)(x+0.430964)(x+1.54863)(x+2.65236x)$\\\\
$\zeta=4$&$x^2+ 1.270 x + 0.5564 $ &$-12(x+2.63447)(x+1.4683x)(x-0.430964)(x-1.54863)(x-2.65236x)$\\\\
$\zeta=5$&$(x-2.582)(x-0.6325)$&$-12(x+0.4792)(x+1.5636)(x+2.6585)(x^2- 3.3621 x +2.9644)$\\\\
$\zeta=6$&$(x+2.582)(x+0.6325)$&$-12(x-0.4792)(x-1.5636)(x-2.6585)(x^2+ 3.3621 x +2.9644)$\\\\
$\zeta=7$&$(x-2.620)(x+0.4611)$&$-12(x-0.779279)(x+1.5569)(x+2.6556)(x^2- 2.5336 x +1.66393)$\\\\
$\zeta=8$&$(x+2.620)(x-0.4611)$&$-12(x+0.779279)(x-1.5569)(x-2.6556)(x^2+ 2.5336 x +1.66393)$\\\\
$\zeta=9$&$(x-1.484)(x+0.4196)$&$-12 (x-2.63681)(x+1.5462)(x+2.6515)(x^2- 1.1171 x + 0.4745)$\\\\
$\zeta=10$&$(x+1.484)(x-0.4196)$&$-12 (x+2.63681)(x-1.5462)(x-2.6515)(x^2+ 1.1171 x + 0.4745)$\\\\
$\zeta=11$&$x^2 - 3.831 x + 3.719$&$-12 (x-2.5078)(x-0.6085)(x+0.4859) (x+1.5666)(x+2.6598)$\\\\
$\zeta=12$&$x^2 + 3.831 x + 3.719$&$-12 (x+2.5078)(x+0.6085)(x-0.4859) (x-1.5666)(x-2.6598)$\\\\
$\zeta=13$&$(x-2.637)(x+1.546)$&$-12(x-1.4827)(x+0.4211)(x+2.6516)(x^2- 1.1358 x+ 0.4844)$\\\\
$\zeta=14$&$(x+2.637)(x-1.546)$&$-12(x+1.4827)(x-0.4211)(x-2.6516)(x^2+ 1.1358 x+ 0.4844)$\\
\hline\hline
\end{tabular}
\end{table*}

\begin{table*}[htb]
\caption{The extended  Heine-Stieltjes Polynomials $y^{(\zeta)}_{3}(x)$ for constructing $S=1/2$ states with $n=7$ and $t=3$
according to (35) and the corresponding Van Vleck Polynomials $V^{(\zeta)}_{5}(x)$.}
\begin{tabular}{ccc}
\hline\hline
&{The extended  Heine-Stieltjes Polynomials $y^{(\zeta)}_{3}(x)$} &{The Van Vleck Polynomials $V^{(\zeta)}_{5}(x)$} \\
\hline
\hline
$\zeta=1$&$(x-0.6354)(x^2- 2.020 x + 2.296 )$&$-15 (x-2.581) (x-1.432) (x+0.4002) (x+1.514) (x+2.630)$\\\\
$\zeta=2$&$(x+0.6354)(x^2+ 2.020 x + 2.296 )$ &$-15 (x+2.581) (x+1.432) (x-0.4002) (x-1.514)(x-2.630)$\\\\
$\zeta=3$&$(x-2.542) (x^2+ 0.5857 x +0.2296)$&$-15(x-0.4874) (x+1.473) (x+2.619) (x^2 - 3.213 x + 2.754)$\\\\
$\zeta=4$&$(x+2.542) (x^2- 0.5857 x +0.2296)$&$-15(x+0.4874) (x-1.473) (x-2.619) (x^2 + 3.213 x + 2.754)$\\\\
$\zeta=5$&$(x-2.582) (x+0.5705) (x+2.540)$&$-15(x-0.6329) (x^2- 2.867 x +2.204 ) (x^2+3.394 x + 3.026 )$\\\\
$\zeta=6$&$(x+2.582) (x-0.5705) (x-2.540)$&$-15(x+0.6329) (x^2+ 2.867 x +2.204 ) (x^2-3.394 x + 3.026 )$\\\\
$\zeta=7$&$(x-0.3106) (x^2+ 2.634 x +1.912)$&$-15(x-2.625) (x-1.498) (x+2.562)(x^2+ 1.097 x +0.4387)$\\\\
$\zeta=8$&$(x+0.3106) (x^2- 2.634 x +1.912)$&$-15(x+2.625) (x+1.498) (x-2.562)(x^2- 1.097 x +0.4387)$\\\\
$\zeta=9$&$(x-1.434)(x+0.5286)(x+2.529)$&$-15(x-2.612)(x^2 - 1.116 x +0.4798) (x^2+3.403 x +3.044)$\\\\
$\zeta=10$&$(x+1.434)(x-0.5286)(x-2.529)$&$-15(x+2.612)(x^2 + 1.116 x +0.4798) (x^2-3.403 x +3.044)$\\\\
$\zeta=11$&$(x-2.593) (x^2+ 3.468 x + 3.144 )$&$-15(x-0.7007)(x+0.5593)(x+2.528)(x^2- 2.562 x +1.738)$\\\\
$\zeta=12$&$(x+2.593) (x^2- 3.468 x + 3.144 )$&$-15(x+0.7007)(x-0.5593)(x-2.528)(x^2+ 2.562 x +1.738)$\\\\
$\zeta=13$&$(x-1.461) (x^2 + 3.299 x +2.884)$&$-15(x-2.616)(x+0.5086)(x+2.537)(x^2- 0.7961 x +0.3266 )$\\\\
$\zeta=14$&$(x+1.461) (x^2 - 3.299 x +2.884)$&$-15(x+2.616)(x-0.5086)(x-2.537)(x^2+ 0.7961 x +0.3266 )$\\
\hline\hline
\end{tabular}
\end{table*}

\section{V. Application to Identical-Particle Systems}

Classification and construction of  identical-particle states
for a given angular momentum quantum number are fundamental, especially
in nuclear structure theory. $n$-coupled states of $l$-bosons
can be constructed as the basis vectors of symmetric irreducible representations
of $U(2l+1)\supset O(2l+1)\supset O(3)$
as shown in \cite{ham, iac, cap}, while those of $j$-fermions
can be constructed as the basis vectors of antisymmetric irreducible representations
of $U(2j+1)\supset Sp(2j+1)\supset O(3)$ as shown in \cite{flo,tal}.
The Bethe ansatz method for angular momentum projection
with the Heine-Stieltjes correspondence shown in previous sections
can also be applied to construct {states} with definite angular momentum
from a set of uncoupled single-particle product states for
identical-particle systems, which can be done as follows:
Firstly, we solve the BAEs (\ref{4}) for non-identical particle systems with the same spin, and
then to construct the coupled state (\ref{2}). Once the coupled state (\ref{2}) is
expanded in terms of  single-particle product states, we take all
particles to be identical, {which
is} called assimilation. For identical-fermion systems,
the Pauli principle forbidden single-particle product states
will be automatically ruled out after the assimilation.
Because of the additional permutation symmetry
with respect to the single-particle coordinate permutations,
the procedure outlined in previous sections can be simplified.
In this section, we will show how the the procedure is taken.

\vskip .7cm
\noindent {\bf (1) Identical bosons}
\vskip .2cm
Let the single-particle states of boson with angular momentum $l$
be $\vert l,m\rangle\equiv\vert m\rangle$ with $m=-l,-l+1,\cdots,l$. According to (\ref{2}),
$n$-coupled state with total angular momentum $L=nl-k$  and $M_{L}=L$

\begin{equation}\label{bbs}
\vert \zeta; L,M_L=L)=L_{-}(x^{(\zeta)}_{1})\cdots L_{-}(x^{(\zeta)}_{k})\vert{\rm h.w.}\rangle,
\end{equation}
where
$\vert{\rm h.w.}\rangle=\prod_{\alpha=1}^{n}\vert m_{\alpha}= l\rangle$
is the highest weight state,

\begin{equation}
L_{-}(x^{(\zeta)}_{i})=\sum_{\alpha=1}^{n}{1\over{x^{(\zeta)}_{i}-\epsilon_{\alpha}}}L_{-}^{\alpha},
\end{equation}
in which the parameters $\{\epsilon_{\alpha}\}$
can usually be any set of unequal numbers, and $L_{-}^{\alpha}$ is the angular momentum lowering
operator only acting on the $\alpha$-th copy of single-particle state, and
$L_{+}=\sum_{\alpha}L_{+}^{\alpha}$, similar to the non-identical
particle case. The corresponding BAEs is

\begin{equation}\label{bbae}
\sum_{\alpha=1}^{n}{2l\over{x^{(\zeta)}_{i}-\epsilon_{\alpha}}}
-\sum_{t=1(\neq i)}^{k}{2\over{x^{(\zeta)}_{i}-x^{(\zeta)}_{t}}}=0
\end{equation}
for $i=1,2,\cdots,k$.
By substituting the solutions $\{x_{i}\}$ of (\ref{bbae}) into (\ref{bbs}),
(\ref{bbs}) gives final result after assimilation.

It can be easily proven that the $n$-coupled state with $L=ln-1$ is
zero. Because

$$
L^{\alpha}_{-}\vert{\rm h.w.}\rangle=
\sqrt{2l}
\prod_{\beta=1(\neq\alpha)}^{n}\vert m_{\beta}= l\rangle\vert m_{\alpha}=l-1\rangle$$
\begin{equation}
=\sqrt{2l}\prod_{\beta=1}^{n-1}\vert m_{\beta}= l\rangle
\vert m_{n}=l-1\rangle
\end{equation}
 due to the fact that these bosons are identical,
(\ref{bbs}) becomes

$$
\vert \zeta; L=M_L=nl-1)=L_{-}(x^{(\zeta)})\vert{\rm h.w.}\rangle=
$$
\begin{equation}
\sqrt{2l}\sum_{\alpha=1}^{n}{2l\over{x^{(\zeta)}-\epsilon_{\alpha}}}
\prod_{\beta=1}^{n-1}\vert  m_{\beta}= l\rangle\vert m_{n}=l-1\rangle
\end{equation}
which is zero because

\begin{equation}
\sum_{\alpha=1}^{n}{2l\over{x^{(\zeta)}-\epsilon_{\alpha}}}=0
\end{equation}
according to Eq. (\ref{bbae}) when $k=1$.

When $k\geq 2$, the number of states (\ref{bbs}) with $L=2l-k$
may be calculated in the following way:
Let $P_{n}(k)$ be number of different $n$-partitions of
the integer $k$ with $k=\sum_{i=1}^{n}\xi_{i}$, where
$2l\geq \xi_{1}\geq\xi_{2}\geq\cdots\geq\xi_{n}\geq 0$.
Then, the number of linearly independent states shown in (\ref{bbs})
for $l$-bosons $D_{\rm B}(n,k)
=P_{n}(k)-P_{n}(k-1)$, which gives the multiplicity of given $L=nl-k$
in the reduction $U(2l+1)\downarrow O(3)$ for the symmetric irreducible
representation $[n,\dot{0}]$ of $U(2l+1)$. Generally, $D_{\rm B}(n,k)$ is far
less than $d(n,k)$ shown in (6) for non-identical particles.
Therefore, for given $L$, $n$-coupled states (\ref{bbs}) obtained from solutions of (\ref{bbae})
are over-complete for identical-particle systems when $k\geq 2$.
Actually,  (\ref{bbs}) obtained
from different solutions of (\ref{bbae}), up to a normalization constant, are all
the same when $D_{\rm B}(n,k)=1$.
When  $D_{\rm B}(n,k)\geq2$, the solutions (\ref{bbs})
are not orthogonal with respect to the multiplicity label, and many solutions of (\ref{bbs})
can be expressed by a linear combination of other solutions
of (\ref{bbs}).

Simplification can be made to overcome such complexity mainly
because there is a freedom to choose the parameters $\{\epsilon_{\alpha}\}$
in (44). When $D_{\rm B}(n,k)=1$, we set

\begin{equation}
\left\{
\begin{tabular}{c}
$\epsilon_{\alpha}=-1~{\rm for}~\alpha\leq p$, \\
$\epsilon_{\alpha+p}=1~~{\rm for}~\alpha\geq1$,~\\
\end{tabular}\right.
\end{equation}
when $n=2p$, and

\begin{equation}
\left\{
\begin{tabular}{c}
$\epsilon_{\alpha}=-1~\alpha\leq p+1$,~~~~ \\
$\epsilon_{\alpha+p+1}=1~~{\rm for}~\alpha\geq1$,\\
\end{tabular}\right.
\end{equation}
when $n=2p+1$. With such choice, Eq. (\ref{bbae}) becomes

\begin{equation}\label{bbae1}
{2l(p+r)\over{x_{i}+1}}+{2lp\over{x_{i}-1}}
-\sum_{t=1(\neq i)}^{k}{2\over{x_{i}-x_{t}}}=0
\end{equation}
for $i=1,2,\cdots,k$, where $r=0$ when $n=2p$ and $r=1$ when $n=2p+1$, which
are exactly the Niven equations for zeros of the Jacobi polynomial
$P_{k}^{[-2lp-1,-2l(p+r)-1]}(x)$. There is only one set of zeros
of (\ref{bbae1}) which is sufficient for (\ref{bbs}) when
$D_{\rm B}(n,k)=1$. Therefore, (\ref{bbs}) with zeros of
the Jacobi polynomial $P_{k}^{[-2lp-1,-2l(p+r)-1]}(x)$
are $n$-coupled states with $L=nl-k$ when the parameters
$\{\epsilon_{\alpha}\}$ are chosen according to (49) or (50)
when $D_{\rm B}(n,k)=1$.

For example, there is only one state with $L=6$ for  $n=4$ $d$-bosons
($l=2$).
According to (49), we set $\{\epsilon_{1}=\epsilon_{2}=-1,
\epsilon_{3}=\epsilon_{4}=1\}$. Substituting two zeros
$\{x_{1}=-0.2582\imath, x_{2}=0.2582\imath\}$
of the Jacobi polynomial $P_{2}^{[-7, -7]}(x)$ into (\ref{bbs}), we get

\begin{equation}
\vert L=M_L=6\rangle=-0.5222\vert 2, 2, 1, 1\rangle+
0.8528\vert 2, 2, 2, 0\rangle
\end{equation}
after assimilation and normalization.

When $D_{\rm B}(n,k)\geq2$, we have many ways to set
the parameters $\{\epsilon_{\alpha}\}$. The simplest way
is to choose the two-value parameterization with
$\epsilon_{\alpha_{1}}=\epsilon_{\alpha_{2}}=\cdots=\epsilon_{\alpha_{r}}=-1$
and the rest parameters
$\epsilon_{\beta}=1$ when $\beta\neq\alpha_{i}$
for $i=1,2,\cdots,r$. Obviously, there are $2^{n}-2$
different ways of such parameterization, from which
one can choose  $D_{\rm B}(n,k)$ of them. Zeros of
the corresponding Jacobi polynomial can be used
to obtain the final results from (\ref{bbs}).
It seems that $D_{\rm B}(n,k)\leq 2^{n}-2$ is
always satisfied for $n\geq 2$ though we are unable to prove this
inequality in general. Therefore,
the above two-value parameterization seems
sufficient to resolve the multiplicity.

For example, there are two coupled states of $n=4$ $d$-bosons with $L=4$.
One can set $\{\epsilon_{1}=\epsilon_{2}=-1,
\epsilon_{3}=\epsilon_{4}=1\}$ for one solution with

$$\vert\zeta=1,L=M_L=4\rangle= 0.2208\vert 1, 1, 1, 1\rangle -$$$$
0.7211\vert 2, 1, 1, 0\rangle +
0.6403\vert 2, 2, 0, 0\rangle -$$
\begin{equation}
0.1030\vert 2, 2, 1, -1\rangle +0.1030\vert 2, 2, 2, -2\rangle\end{equation}
and set $\{\epsilon_{1}=-1,\epsilon_{2}=
\epsilon_{3}=\epsilon_{4}=1\}$ for another solution with

$$\vert\zeta=2,L=M_L=4\rangle= 0.0827\vert 1, 1, 1, 1\rangle -$$$$
0.2702 \vert 2, 1, 1, 0\rangle -
0.1161\vert 2, 2, 0, 0\rangle +$$
\begin{equation}0.6733\vert 2, 2, 1, -1\rangle -0.6733\vert 2, 2, 2, -2\rangle.\end{equation}

In this case, the final coupled states (53) and (54) are  not orthogonal
with respect to the multiplicity label $\zeta$, namely
$\langle\zeta=1\vert\zeta=2\rangle\neq0$.
In order to be orthonormalized, the Gram-Schimidt
process may be adopted.

More complicated parameterizations are always possible.
For example, we can also set

\begin{equation}
\left\{
\begin{tabular}{c}
$\epsilon_{\alpha}=-1~{\rm for}~\alpha\leq p$,~~~~~ \\
$\epsilon_{p+1}=0$,~~~~~~~~~~~~~~~~~~ \\
$\epsilon_{\beta+p+1}=1~~{\rm for}~\beta\geq1$,\\
\end{tabular}\right.
\end{equation}
where the integer $p$ can arbitrarily be chosen.
Thus, the BAEs (\ref{bbae}) become

\begin{equation}\label{bbae2}
{2lp\over{x^{(\zeta)}_{i}+1}}+{2l\over{x^{(\zeta)}_{i}}}+{2l(n-p-1)\over{x^{(\zeta)}_{i}-1}}
-\sum_{t=1(\neq i)}^{k}{2\over{x^{(\zeta)}_{i}-x^{(\zeta)}_{t}}}=0
\end{equation}
for $i=1,2,\cdots,k$.
Then, one can choose $D_{\rm B}(n,k)$ solutions of (\ref{bbae2})
to get the results. When we set
$\{\epsilon_{1}=-1,\epsilon_{2}=0,
\epsilon_{3}=\epsilon_{4}=1\}$ for the previous $L=4$ example
of $4$ $d$-bosons, there are $5$ solutions of (\ref{bbae2})
with the extended Heine-Stieltjes polynomials shown in Table VI.

\begin{table}[htb]
\caption{The extended Heine-Stieltjes Polynomials $y^{(\zeta)}_{4}(x)$ for $n=4$ $d$-bosons coupled to $L=M_L=4$
with $\{\epsilon_{1}=-1,\epsilon_{2}=0, \epsilon_{3}=\epsilon_{4}=1\}$.}
\begin{tabular}{cc}
\hline\hline
&{The extended  Heine-Stieltjes Polynomials $y^{(\zeta)}_{4}(x)$} \\
\hline
\hline\\
$\zeta=1$&$(0.1382-0.7053x+x^2)(0.2800-0.6172x + x^2)$\\
\\
$\zeta=2$&$(0.6106+1.5493x+x^2)(0.9545+1.8343x + x^2)$\\\\
$\zeta=3$&$(-0.2918+x)(0.4284+x)(0.1284-0.4857x + x^2)$\\\\
$\zeta=4$&$ (0.0876-0.5298x +x^2)(0.3138+1.0483 x + x^2)$\\\\
$\zeta=5$&$(-0.2874+x) (0.6446+x)(0.6007+1.4124x + x^2)$\\\\
\hline\hline
\end{tabular}
\end{table}

The corresponding coupled states after normalization are

$$\vert\zeta=1,L=M_L=4\rangle=
0.1974\vert 1, 1, 1, 1\rangle-$$
$$0.6448\vert 2, 1, 1, 0\rangle+0.6502\vert 2, 2, 0, 0\rangle-$$
$$0.2475\vert 2, 2, 1, -1\rangle+ 0.2475\vert 2, 2, 2, -2\rangle;$$

$$\vert\zeta=2,L=M_L=4\rangle=
-0.0432\vert 1, 1, 1, 1\rangle+$$
$$0.1412\vert 2, 1, 1, 0\rangle+0.2252\vert 2, 2, 0, 0\rangle-$$
$$0.6810\vert 2, 2, 1, -1\rangle+ 0.6810\vert 2, 2, 2, -2\rangle;$$

$$\vert\zeta=3,L=M_L=4\rangle=
0.1926\vert 1, 1, 1, 1\rangle-$$
$$0.6290\vert 2, 1, 1, 0\rangle+0.2642\vert 2, 2, 0, 0\rangle+$$
$$0.4987\vert 2, 2, 1, -1\rangle- 0.4987\vert 2, 2, 2, -2\rangle;$$

$$\vert\zeta=4,L=M_L=4\rangle=
0.0829\vert 1, 1, 1, 1\rangle-$$
$$0.2707\vert 2, 1, 1, 0\rangle+0.5085\vert 2, 2, 0, 0\rangle-$$
$$0.5750\vert 2, 2, 1, -1\rangle+0.5750\vert 2, 2, 2, -2\rangle;$$

$$\vert\zeta=5,L=M_L=4\rangle=-
0.2355\vert 1, 1, 1, 1\rangle+$$
$$0.7692\vert 2, 1, 1, 0\rangle-0.5740\vert 2, 2, 0, 0\rangle-$$
$$0.1081\vert 2, 2, 1, -1\rangle+0.1081\vert 2, 2, 2, -2\rangle.$$

Because $D_{\rm B}(4,4)=2$ in this case, we may choose

$$\vert\chi=1\rangle=\vert\zeta=1\rangle,$$

$$\vert\chi=2\rangle=c_{1}\vert\zeta=1\rangle+c_{2}\vert\zeta=2\rangle,$$
where $c_1=1/{\cal N}$ and $c_2=-{1\over{{\cal N}\langle\zeta=1\vert\zeta=2\rangle}}$
with the normalization constant

$${\cal N}=\left( \langle\zeta=1\vert\zeta=1\rangle+
{\langle\zeta=2\vert\zeta=2\rangle\over{\langle\zeta=1\vert\zeta=2\rangle^2}}-2\right)^{1/2}$$
according to the Gram-Schmidt process.
Then, one finds

$$\vert\zeta=3\rangle=0.3686\vert\chi=1\rangle+0.9296\vert\chi=2\rangle,$$

$$\vert\zeta=4\rangle=0.8062\vert\chi=1\rangle-0.59168\vert\chi=2\rangle,$$

$$\vert\zeta=5\rangle=-0.8622\vert\chi=1\rangle-0.5066\vert\chi=2\rangle.$$
This example shows that coupled states with zeros of other polynomials
can indeed be expressed as linear combinations of the chosen two due to
the overcompleteness.

\vskip .3cm
\noindent{\bf (2) Identical fermions}
\vskip .3cm
Let the single-particle states of fermions with spin $j$
be $\vert j,m\rangle\equiv\vert m\rangle$ with $m=-j,-j+1,\cdots,j$.
Unlike identical bosons, we have verified that the parameters
$\{\epsilon_{\alpha}\}$ must be a set of unequal numbers for identical fermions.
Initially, we need to solve BAEs (\ref{4}) for non-identical particles
with the same spin $j_{\alpha}=j$ for $\alpha=1,2,\cdots,n$.
After (\ref{2})  is expanded in terms of the single-particle
product states, we then take all particles to be identical.
The Pauli exclusion will automatically rule out any forbidden
single-particle product states after such assimilation.
The  result of (\ref{2})
gives final coupled state with total angular momentum $J=nj-k$  and $M_{J}=J$.
However, the two-value parameterization schemes for identical
bosons shown previously can not be used for identical fermions
mainly because the single-particle product states are totally antisymmetric
with respect to permutations among different single-particle states.
As a consequence, the coupled state is zero if one choose any
two-value parameterization scheme in $\{\epsilon_{\alpha}\}$ for identical fermions.

Similar to identical bosons, the number of linearly independent
states obtained from (\ref{2}), $D_{\rm F}(n,k)$, can be calculated
as follows: Let $Q_{n}(k)$ be number of different $n$-partitions of
the integer $k$ with $k=\sum_{i=1}^{n}\xi_{i}$, where
$2j+1-n\geq \xi_{1}\geq\xi_{2}\geq\cdots\geq\xi_{n}\geq 0$.
Then, the number of linearly independent states obtained from (\ref{2})
for $j$-fermions, $D_{\rm F}(n,k)
=Q_{n}(k)-Q_{n}(k-1)$, which gives the multiplicity of given $J=nj-k$
in the reduction $U(2j+1)\downarrow O(3)$ for the antisymmetric irreducible
representation $[1^{n},\dot{0}]$ of $U(2l+1)$. Generally, $D_{\rm F}(n,k)$ is far
less than $d(n,k)$ shown in (6) for non-identical particles.
Therefore, (\ref{2})  obtained from solutions of (\ref{4}) are
also over-complete.
Similar to identical bosons, one only needs to choose $D_{\rm F}(n,k)$
solutions of (\ref{4}). When the parameters $\{\epsilon_{\alpha}\}$
are chosen according to (29) or (30), the coupled state
(\ref{2}) satisfies the symmetry

$$J_{-}(x^{(\zeta)}_{1})J_{-}(x^{(\zeta)}_{2})\cdots J_{-}(x^{(\zeta)}_{k})\vert{\rm h.w.}\rangle$$
\begin{equation}
=J_{-}(-x^{(\zeta)}_{1})J_{-}(-x^{(\zeta)}_{2})\cdots J_{-}(-x^{(\zeta)}_{k})\vert{\rm h.w.}\rangle.
\end{equation}
Therefore, only one of reflectional symmetry pair of Stieltjes zeros
should be considered.

In the following, we take $n=3$ $j=9/2$ identical fermions as examples.
In this case, there is only one coupled state with $J=M_J=17/2$ ($k=5$), for
which there are $6$ extended Heine-Stieltjes polynomials as shown in Table VII,
which clearly shows that  $y^{(2)}_{5}(x)=y^{(1)}_{5}(-x)$,
$y^{(4)}_{5}(x)=y^{(3)}_{5}(-x)$, and
$y^{(6)}_{5}(x)=y^{(5)}_{5}(-x)$.
But three solutions  $y^{(1)}_{5}(x)$, $y^{(3)}_{5}(x)$,
and $y^{(5)}_{5}(x)$ all result in one coupled state

$$
\vert J=M_J=17/2\rangle=
0.7746~ \vert 9/2, 5/2, 3/2\rangle -$$
\begin{equation}
 0.6325~ \vert 9/2, 7/2, 1/2\rangle
\end{equation}
up to a normalization constant after assimilation.

\begin{table}[htb]
\caption{The extended Heine-Stieltjes Polynomials $y^{(\zeta)}_{5}(x)$ for  $J=M_J=17/2$ ($k=5$) coupled states of $n=3$ $j=9/2$ identical
fermions with $\{\epsilon_{1}=-1, \epsilon_{2}=0, \epsilon_{3}=1\}$.}
\begin{tabular}{cc}
\hline\hline
&{The extended  Heine-Stieltjes Polynomials $y^{(\zeta)}_{5}(x)$} \\
\hline
\hline\\
$\zeta=1$&$ (-0.6191+x) (0.60249-1.3616 x+x^2)\times$\\
&$(0.4249-1.2622x+x^2)$\\\\
$\zeta=2$&$ (0.6191+x) (0.60249+1.3616 x+x^2)\times$\\
&$(0.4249+1.2622x+x^2)$\\\\
$\zeta=3$&$(0.4849+x) (0.4559-1.2046x+x^2)\times$\\
&$(0.34201-1.1556x+x^2)$\\\\
$\zeta=4$&$(-0.4849+x) (0.4559+1.2046x+x^2)\times$\\
&$(0.34201+1.1556x+x^2)$\\\\
$\zeta=5$&$(0.5371+x) (0.2723-1.0146x+x^2)\times $\\
&$(0.3451-1.0889x+x^2)$\\\\
$\zeta=6$&$(-0.5371+x) (0.2723+1.0146x+x^2)\times $\\
&$(0.3451+1.0889x+x^2)$\\
\hline\hline
\end{tabular}
\end{table}

\begin{table}[htb]
\caption{The extended Heine-Stieltjes Polynomials $y^{(\zeta)}_{9}(x)$ for  $J=9/2$ coupled states of $n=3$ $j=9/2$ identical
fermions with $\{\epsilon_{1}=-1, \epsilon_{2}=0, \epsilon_{3}=1\}$.}
\begin{tabular}{cc}
\hline\hline
&{The extended  Heine-Stieltjes Polynomials $y^{(\zeta)}_{9}(x)$} \\
\hline
\hline\\
$\zeta=1$&$ (1.7186-2.3241 x+x^2)(0.9532-1.7796x+x^2)\times$\\
&$(0.6707-1.5615x+x^2) (0.5516-1.4664x+x^2)\times$\\
&$(-0.7196+x)$\\\\
$\zeta=2$&$ (1.7186+2.3241 x+x^2)(0.9532+1.7796x+x^2) \times$\\
&$(0.6707+1.5615x+x^2) (0.5516+1.4664x+x^2)\times$\\
&$(0.7196+x)$\\\\
$\zeta=3$&$ (1.3423-1.8543 x+x^2)(0.7134-1.4807 x+x^2)\times$\\
&$(0.4992-1.3433 x+x^2)(0.4228-1.2929 x+x^2)\times$\\
&$(0.4142+x)$\\\\
$\zeta=4$&$ (1.3423+1.8543 x+x^2)(0.7134+1.4807 x+x^2)\times$\\
&$(0.4992+1.3433 x+x^2)(0.4228+1.2929 x+x^2)\times$\\
&$(-0.4142+x)$\\\\
$\zeta=5$&$ (0.9728-1.4201x+x^2)(0.5034-1.2196x+x^2)\times$\\
&$(0.3611-1.1571x+x^2)(0.1818+0.8182x+x^2)\times$\\
&$(-0.5708+x)$\\\\
$\zeta=6$&$(0.9728+1.4201x+x^2)(0.5034+1.2196x+x^2)\times$\\
&$(0.3611+1.1571x+x^2)(0.1818-0.8182x+x^2)\times$\\
&$(0.5708+x) $\\\\
$\zeta=7$&$(0.6395-1.0463x+x^2)(0.3362-1.0087x+x^2)\times$\\
&$(0.2589-1.0027x+x^2)(0.2069+0.7932x+x^2)\times$\\
&$(0.4142+x) $\\
$\zeta=8$&$(0.6395+1.0463x+x^2)(0.3362+1.0087x+x^2)\times$\\
&$(0.2589+1.0027x+x^2)(0.2069-0.7932x+x^2)\times$\\
&$(-0.4142+x) $\\
$\zeta=9$&$ (0.1801-0.8267x+x^2)(0.2527-0.7578x+x^2)\times$\\
&$(0.3775+0.7973x+x^2)(0.2253+0.8689x+x^2)\times$\\
&$(0.4423+x)$\\
$\zeta=10$&$ (0.1801+0.8267x+x^2)(0.2527+0.7578x+x^2)\times$\\
&$(0.3775-0.7973x+x^2)(0.2253-0.8689x+x^2)\times$\\
&$(-0.4423+x)$\\
\hline\hline
\end{tabular}
\end{table}

Since $D_{\rm F}(3,9)=2$ for $j=9/2$ identical fermions, $J=9/2$ should occur twice.
While there are $10$ extended Heine-Stieltjes polynomials as shown in Table VIII, which
shows that  $y^{(2)}_{9}(x)=y^{(1)}_{9}(-x)$,
$y^{(4)}_{9}(x)=y^{(3)}_{9}(-x)$,
$y^{(6)}_{9}(x)=y^{(5)}_{9}(-x)$,
$y^{(8)}_{9}(x)=y^{(7)}_{9}(-x)$,
and $y^{(10)}_{9}(x)=y^{(9)}_{9}(-x)$,
we only need to choose $2$ of them to get the coupled states
according to (\ref{2}). The coupled state with $y^{(1)}_{9}(x)$
is

$$\vert\zeta=1, J=M_J=9/2\rangle=0.2105~\vert 5/2, 3/2, 1/2\rangle -$$$$
 0.1684~\vert 7/2, 3/2, -1/2\rangle +
 0.1575~\vert 7/2, 5/2, -3/2\rangle-$$$$
 0.3384~\vert 9/2, 1/2, -1/2\rangle +
 0.4415~ \vert 9/2, 3/2, -3/2\rangle -$$$$
 0.5446~\vert 9/2, 5/2, -5/2\rangle +
 0.5446~\vert 9/2, 7/2, -7/2\rangle,$$
and that with $y^{(3)}_{9}(x)$ is

$$\vert\zeta=2, J=M_J=9/2\rangle=
0.1506~\vert 5/2, 3/2, 1/2\rangle -$$$$
0.1205~\vert 7/2, 3/2, -1/2\rangle +
0.1127~\vert 7/2, 5/2, -3/2\rangle -$$$$
0.3913~\vert 9/2, 1/2, -1/2\rangle +
0.4651~\vert 9/2, 3/2, -3/2\rangle -$$$$
0.5388~\vert 9/2, 5/2, -5/2\rangle +
0.5388~\vert 9/2, 7/2, -7/2\rangle.$$

After the Gram-Schmidt orthonormalization, we have

$$\vert\chi=1, J=M_J=9/2\rangle=\vert\zeta=1, J=M=9/2\rangle,$$

$$\vert\chi=2, J=M_J=9/2\rangle=
0.5526\vert 5/2, 3/2, 1/2\rangle -$$$$
0.4421\vert 7/2, 3/2, -1/2\rangle +
0.4135\vert 7/2, 5/2, -3/2\rangle +$$$$
0.5164\vert 9/2, 1/2, -1/2\rangle -
0.2457\vert 9/2, 3/2, -3/2\rangle -$$$$
0.0250\vert 9/2, 5/2, -5/2\rangle +
0.0250\vert 9/2, 7/2, -7/2\rangle.$$
Then, other $3$ coupled states corresponding
to $y^{(5)}_{9}(x)$, $y^{(7)}_{9}(x)$, $y^{(9)}_{9}(x)$, respectively,
can be expressed as linear combinations of $\vert\chi=1\rangle$
and  $\vert\chi=2\rangle$.

\section{VI. Summary}
In summary, a new angular momentum projection for many-particle systems
is formulated based on the Heine-Stieltjes correspondence,
which can be regarded as the solutions of the mean-field plus
the paring model in the strong pairing interaction $G\rightarrow\infty$ limit~\cite{Guan}.
With the special choice of the parameters $\{\epsilon_{\alpha}\}$,
the solutions of the associated BAEs are simplified because of
the additional reflection symmetries. Properties of the Stieltjes
zeros and the related Van Vleck zeros are discussed.
The electrostatic interpretation of these zeros are presented.
As an example, the application to $n$ nonidentical particles with spin-$1/2$
is made to elucidate the procedure and properties of the Stieltjes
zeros and the related Van Vleck zeros. It is clear that the new angular momentum
projection can be used for nonidentical-particles with arbitrary spins.
It is shown that the new angular momentum projection for identical bosons
or fermions can be simplified with the branching multiplicity formula
of $U(N)\downarrow O(3)$ and the special choices of the parameters used
in the projection. Especially, it is shown that the coupled states of identical
bosons can always be expressed in terms of zeros of Jacobi polynomials.
However, unlike non-identical particle systems,
the coupled states of identical particles are non-orthogonal
with respect to the multiplicity label after the projection.
In order to establish orthonormalized coupled states for identical particles,
the Gram-Schimidt process may be adopted.
It will be advantageous in the application, for example, to the shell model calculations
if matrix elements of one- and two-body operators
under the angular momentum projected basis can be calculated easily, which
seems possible as shown in~\cite{rg2} where explicit expressions for the
expectation values of one- and two-body operators in the mean-field plus pairing
model were obtained, of which the relevant research is in progress.

\begin{acknowledgments}
One of the authors (PF) is grateful to Faculty of Science, The University of
Queensland, for support through an Ethel Raybould Visiting Fellowship.
Support from U.S. National Science Foundation (OCI-0904874),
Southeastern Universities Research Association,
Natural Science Foundation of China (11175078), Australian Research Council
(DP110103434), Doctoral Program Foundation of the State Education Ministry
of China (20102136110002), and LSU--LNNU joint research program (9961) is acknowledged.
\end{acknowledgments}

\end{document}